# Optical detection of fractal dimensions of MUC13 stained pancreatic tissues for cancer diagnostics

Prakash Adhikari[1], Aby Panikanthara Binu[2], Shiva Bhandari[3], Sheema Khan[4], Mina Jaggi[4], Shubhash C. Chauhan[4], and Prabhakar Pradhan[1*]

[1]Department of Physics and Astronomy, Mississippi State University, Mississippi State, MS 39762, USA
[2] Rhodes College, Memphis, TN 38112, USA
[3]Department of Physics, East Carolina University, Greenville, NC, USA, 27858
[4]Department of Pharmaceutical Sciences and the Centre for Cancer Research, University of Tennessee Health Science Center, College of Pharmacy, Memphis, TN, USA

*E-mail: PPradhan: pp838@msstate.edu

**Abstract:** Transmission intensity of a thin tissue sample produce intensity distribution that is proportional to the refractive index pattern in the thin slice, in turn reflects mass density pattern, and using the intensity pattern, fractal dimension can be calculated. In this paper, we report fractal dimension analyses of MUC13 stained pancreatic cancer tissues. Pancreatic cancer is deadliest due to its physical orientation, and no prominent chemical change till the late stages of cancer. The presence of pancreatic cancer can be detected using MUC13 dye, however this fails to explain about the stages of cancer. Here we studied MUC13 expressed cancer tissues and observe their cancer stages by fractal dimension of light transmission intensity. Our results show that early stages of pancreatic cancer for MUC13 expressed tissue was accurately distinguished by fractal dimension analysis. Further, its applications to cancer diagnostics is also discussed in detail.

## 1. INTRODUCTION

*Progression of cancer in tissues and fractal dimension:* A tissue is a fractal object due to its immanent self-similar patterns from larger to decreasingly smaller length scales. With the progression of cancer in the tissues, the tissue structure changes which in turn alter the fractality and hence increased in the fractal dimension. As a cancer progresses in tissue, cells inside the tissue changes coherently and they multiply rapidly in asymmetrical pattern resulting to have systematic change in fractal dimensions. This implies a strong correlation between the change in fractal dimension and the progression of cancer.

*Optical transmission intensity of thin tissue samples and fractal dimension detection for cancer.* Optical transmission intensity of thin tissue samples can produce intensity distribution that is proportional to the refractive index pattern in the thin tissue slices, in turn mass density pattern. Using these intensity patterns, mean fractal dimension can be calculated.

*Pancreatic cancer and its difficulties in detection:* The American Cancer Society has reported that in 2018 about 55,440 people will be diagnosed with pancreatic cancer and about 44,330 people will die of pancreatic cancer in the United States. About 3% of all cancers in the US and about 7% of all cancer deaths are reported for pancreatic cancer. Since pancreas is an organ located deep in the stomach i.e. on the right side of abdomen and is connected to the duodenum, it is difficult to diagnose the tumor by physician during routine examination due to its inaccessibility. People with pancreatic cancer develop no symptoms and chemical change even in later stages of cancer. Therefore, accurate detection and staging are important for pancreatic cancer diagnostics and treatments.

The state-of-the-art can detect stages and classify pancreatic cancer through the utilization of fractal dimensional analysis. In this work, the correlation between fractal dimension and pancreatic cancer staging related to MUC13 expressed pancreatic tissue microarray (TMA) samples are reported. The results clarify the pancreatic cancer growth by MUC13 expression in cancerous tissue, and their correlation to fractal dimension.

**1.1 Detection of Pancreatic Cancer**

Pancreatic cancer remains as a major unresolved health problem in this present medical era. As mentioned earlier, American Cancer Society reports that in 2018, pancreatic cancer is about 3% of the total cancers and about 7% of all the cancer deaths in the US. Pancreatic cancer has the worst 5 years survival rate of all the common cancers (Allemani et al., 2018). Since pancreas is deep-seated within the abdomen and pancreatic cancer can spread out from initial to later stages shortly (less than 2 years), it is very critical and important to develop early detection methods. The ongoing histological processes for the detection of cancer are inefficient for the pancreatic cancer detection. Many times, the pathological tests are inconclusive although the patients may have pancreatic cancer symptoms. Very few cases of pancreatic cancer are diagnosed when it is within pancreas, but without tumors. It, therefore, demands for a development of effective diagnostic tool for the early detection of the pancreatic cancer.

**1.2 Pancreatic Cancer and MUC13**

Pancreatic cancer is considered as deadliest due to the lack of reliable early detection methods and underlying mechanisms of its aggressive pathogenesis. Recent studies have shown a newly identified transmembrane mucin, MUC13 plays important role in the diagnostic and therapeutic of pancreatic cancer. MUC13 is a cell surface glycoprotein acts as lubricant and chemical barriers in the cell membrane. The expression profile of MUC13 was found highly significant in cancerous samples compare with normal pancreatic tissues. Therefore, the biomarker MUC13 expressed significantly in pancreatic cancer leaving red sport resulting in higher mass accumulation.

**1.3 Fractal Dimension and quantification**

The development of techniques for the detection of the cancerous cells/tissues is a major challenge in the healthcare industry. However; fractal dimensional analysis can be taken as one of the methods that can overcome this challenge (de Arruda et al., 2013; Gazit, Berk, Leunig, Baxter, & Jain, 1995; Klatt et al., 2014; Landini & Rippin, 1993; Luzi et al., 1999; Metze, 2013; Sankar & Thomas, 2010; Stępień & Stępień, 2010). Fractal dimension analysis essentially serves as a morphometric measure of irregular structures in biological samples, such as those predominantly found in tissues as well as its progression with cancer. It has been reported that the progress of cancer makes tissue more chaotic than the normal tissues (Nagy, Chang, Dvorak, & Dvorak, 2009). Therefore, fractal dimension analysis can act as a tissue diagnostic tool for progressive carcinogenesis. Fractal dimension is a measure of the space-filling capacity of a certain pattern in a designated area, some way porosity of the intensity.

We propose that it is possible to distinguish different stages of pancreatic cancer with the aid of this fractal dimension approach. The experimental study clearly indicates the relationship between a pancreatic cancer cell stage and its fractal dimension. To elaborate, we found variations in the fractal dimension in the two different stages of pancreatic cancer, that is in correlation with the stages of cancer. The aim of this research is to develop a precise method for the early detection of pancreatic cancer cells and stages that will help the medical community to downplay the cancer patients in future. This method may work independently or help in assisting the pathologist.

## 2. METHOD

### 2.1. Fractal Dimension Calculations

*Box-counting methods:* Fractal dimension is a mathematical parameter that exhibits self-similar patterns at different length scales. The fractal structure is assumed to be lying on an evenly spaced grid, and the number of boxes required to cover the fractal structure are counted. In this method, MUC13 dye is added to cancerous tissue then fractal dimension is calculated by observing how the number changes as the grid is refined by applying a box counting algorithm. Therefore, fractal dimension is the measure of space filling capacity of the pattern with the addition of MUC13 stain in cancerous cell (Bhandari, Choudannavar, Avery, Sahay, & Pradhan, 2018). The fractal dimension of the systems follows the scaling power law $N \propto r^{-D_f}$, where $D_f$ being the fractal dimension, and is taken easily enough by the popularized Minkowski–Bouligand dimension, or box counting method, following if an N-dimensional space containing a specific fractal structure is partitioned with a number of N-cubes [$N(r)$] with side length $r$, then the N-cube counting fractal dimension is defined as (Dubuc, Quiniou, Roques-Carmes, Tricot, & Zucker, 1989; Mandelbrot, 1982):

$$D_f = \lim_{r \to 0} \frac{-\ln N(r)}{\ln(r)} \qquad (1)$$

Counting the number of N-cube of different size ($r$), the fractal dimension of equation (4) can be estimated by the slope of the linear fit of [ln (1/r), ln N(r)]. The N-cube sizes were simulated by powers of two with side length $r = 1, 2, 4 \dots 2^W$, where $W$ is the smallest integer such that the length size of the system matrix L is less than or equal to $2^W$. The fractal dimension box counting was calculated by the open source image based NIH software *Image J*.

### 2.2. Pancreatic cancer TMA tissue samples and transmission microscopy image collection

A pancreatic cancer tissue micro array sample TMA (PA241b), ordered from Biomax, has two different stages of pancreatic cancer tissue types. In our study, we used mainly following three categories of pancreatic tissues: stage 1, stage 2 and malignant tumor. The tissue samples were stained with MUC13 stain to see the presence of pancreatic cancer. Therefore, MUC13 dye is stained with pancreatic cancer tissue in detecting and distinguishing the highly affected cancer areas from the unaffected areas. The tissue microarray TMA consists of pancreatic cancer samples with MUC13 stain in the array with diameter 1.5mm for each sample and thickness 5µm. Among all the samples of different stages a total of nine samples i.e. three from each cancer stages as mentioned earlier were considered for imaging. The TMA of pancreatic cancer stained with MUC13 was kept on the sample holding platform of the microscope operating in transmission mode to obtain the optical image. The microscopic images of these samples were captured by Olympus BX50 Microscope (Olympus, USA) and an Infinity2 Microscopy Camera (Lumenera Corp. Canada) attached to the microscope head. Using Lumenera CAPTURE software images at magnification ×40 were sampled and digitized to eight bits of gray level and stored as digital files. Ten set of images for all the regions of the stages of cancer were taken as new data sets to work with. The affected and unaffected regions are distinguished in terms of MUC13 dye expressed in the cells. The regions where the MUC13 dye was expressed were termed as affected regions and unexpressed were termed as unaffected regions. That means MUC13 expression in the tissue indicates the cancerous tissue and results in the higher spatial mass density variation in sample. This mass density variation is linearly related to the transmission intensity in the refractive index in the image. Therefore, fractal dimension of the optical intensity patterns correlates with the mass density pattern.

### 2.3. Fractal analysis from image data

The collected images of the samples were processed by *ImageJ* (NIH, USA) software and analyzed for the fractal dimensions. The fractal dimension analysis tool in *ImageJ* calculated the fractal dimension of each samples, that uses the box counting method. Fractal dimension was calculated for unaffected regions, and affected regions of pancreatic cancer stages. In addition, the average fractal dimension of the overall cell containing both the regions was also calculated for comparisons. The data were organized and processed to calculate the variation in fractal dimension of the different stages of pancreatic cancer along with the net standard deviations and the p-values.

## 3. RESULTS

The fractal dimension of all the stages of pancreatic cancer samples were calculated using box counting method. The observed microscopic images were processed using *imageJ* fractal analysis tool. The captured images of MUC13 unexpressed, average expressed, and expressed regions of stages of pancreatic cancer tissues are shown in Figure 1.

The experiment shows MUC13 dye on pancreatic cancer tissue interact with cancer affected region and remains isolated for normal region. Therefore, in the cancer affected tissue MUC13 dye can express differently in the different region of the pancreatic tissue. The cell or tissue expressed with MUC13 remains as red spot with high mass density fillings, whereas MUC13 unexpressed region has less mass density fillings.

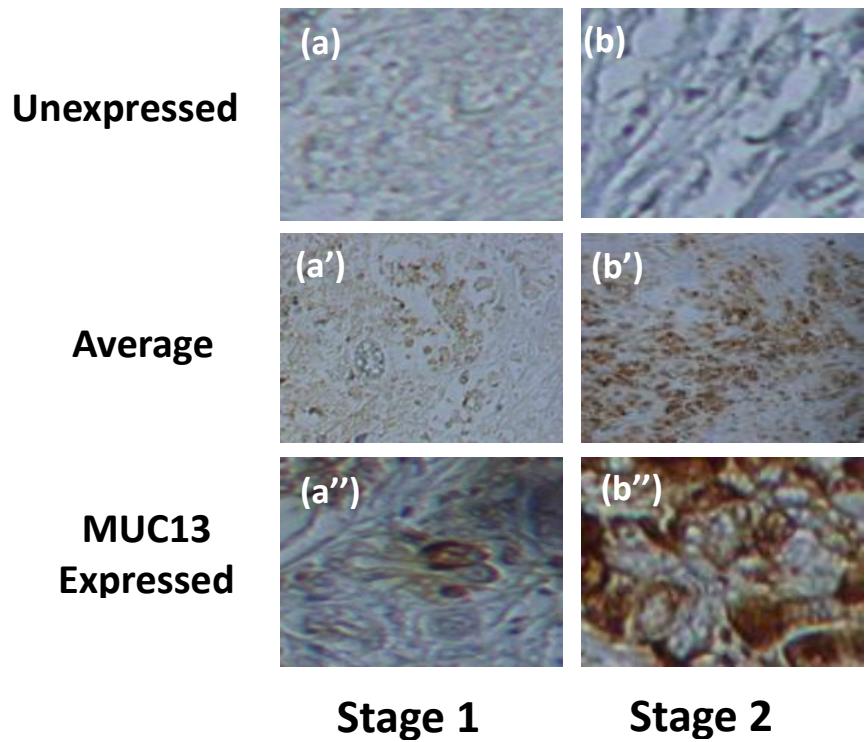

**Figure 1:** Different Stages of pancreatic cancer stained with MUC13 stain. The first row of images **(a-b)** represents MUC13 unexpressed images (optical micrographs), middle row of images **(a'-b')** represents average MUC13 stained images, and bottom row of images **(a''-b'')** represents MUC13 expressed images of stages of pancreatic cancer respectively.

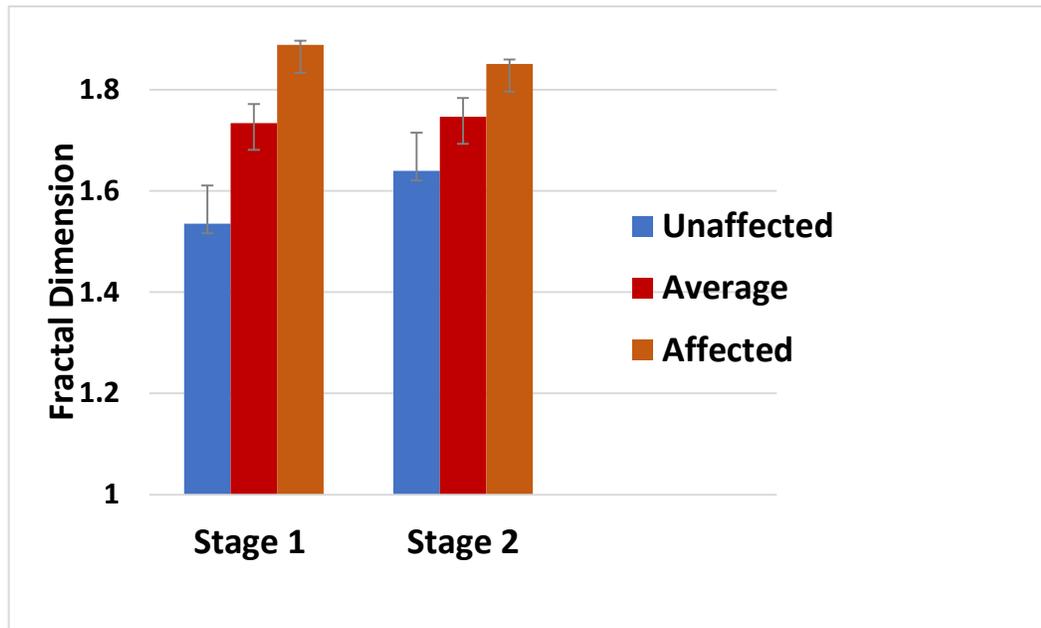

**Figure 2:** Expression of MUC13 stain in different stages of Pancreatic cancer bar graphs**:** This graph shows the distribution of fractal dimension in different stages of pancreatic cancer stained with MUC13 stain. The fractal dimension of the areas expressed with MUC13 is greater than that of unexpressed by MUC13 stain. The average fractal dimension of the cell is also shown which is in between the expressed and the unexpressed areas' values. This trend is similar in both the stages of the pancreatic cancer stained with MUC13 (P-values <0.05).

MUC13 expressed differently in cancer tissue due to this reason, average fractal dimension of the same cancer affected samples were calculated. Fractal dimension was calculated for stages of cancer at three different regions, and an ensemble averaging were performed (over 10 tissues per case). In Figure 2, we presented the bar graph for the fractal dimensions of different cancer stages. As seen, the fractal dimension has a monotonic behavior i.e. saturation for the different stages of the pancreatic cancer that are expressed with the MUC13 dye. In particular, regions with MUC13 unexpressed to MUC13 expressed regions of the samples are shown in Figure 1.

In the bar graphs, the stage 1 pancreatic cancer tissue has the lowest fractal dimension in all the three regions of cancer than the stage 2. The fractal dimension of the region of the cells unaffected by MUC13 dye increases from stage 1 to stage 2 significantly. Therefore, the average fractal dimension of the cell increases the same way. However, the fractal dimension in case of

MUC13 expressed region is not significant and remains almost saturated in both the stages of cancer. The absolute fractal dimension value of the stage 1 and stage 2 cancer for the MUC13 unexpressed regions in the tissues were calculated to be 1.53515, and 1.6397, average expressed regions to be 1.73415, and 1.74615, and expressed regions to be 1.88838, and 1.85105, respectively. These numerical parameters are the logical finding of MUC13 average expressed regions as a strong indicator of detection of early stages of pancreatic cancer for all the regions of observation. The unique and useful aspect of the approach demonstrated MUC13 stain expressed with cancer affected tissues resulting in logical finding of the progress of cancer which is due to more mass accumulation with more space filling in tissue, as mentioned. In particular, this work concluded fractal dimension of TMA samples is the tool to measure the space filling/irregularity of the cell structure that increases as the number of cancerous cells increases.

## 4. CONCLUSIONS AND DISCUSSIONS

This state-of-the-art experiment demonstrated structural properties of pancreatic cancer tissue TMA samples in the presence of MUC13 dye, using transmission microscopy micrographs. As a preliminary result, fractal dimension of different stages of pancreatic cancer were calculated using *ImageJ* software. The numeric index i.e. fractal dimension increases from a normal to a cancerous tissue due to the accumulation of mass density filling with the growth of cancer. Further, the presence of MUC13 stain increases the mass filling of the cancerous cells and hence the fractal dimension increases. And since, MUC13 gets expressed evenly in cancer affected region, the fractal dimension of two different stages of cancer is almost saturated which may not be a perfect indicator of stages of cancer detection. The hierarchy of average value of fractal dimension of MUC13 expressed samples follow stage 1 < stage 2 and is quite significant (higher p-values) than any other regions. Interestingly, the fractal dimension value of MUC13 unexpressed area has the prominent difference. Therefore, MUC13 expressed average cell fractal dimension act as a numeric indicator of stages of pancreatic cancer detection which needs to be explored further.

In Summary, fractal dimension lays foundation for reliable, cost-efficient, easy, and early detection of the pancreatic cancer in future medical world. With this capability, fractal dimension, a purely mathematical too, distinguishes the stages of cancer avoiding the limitations of conventional methods which are highly prone to human error, misguiding, risky, and expensive. Moreover, cancerous tissue with biomarker MUC13 dye was found to more efficient for cancer detection. With further research and clinical study, fractal dimension analysis method can be developed as a reliable diagnostic tool of stages of different cancer diagnosis and can be set as guideline

## Acknowledgments

The work reported here was partially supported by the National Institutes of Health (NIH) grants (R01EB003682 and R01EB016983) and Mississippi State University to Dr. Pradhan.